\documentclass[onecolumn]{emulateapj}
\usepackage{amsmath}

\lefthead{Wada et al.}
\righthead{Molecular Gas around AGNs}

\newcommand{\bea}{\begin{eqnarray} }
\newcommand{\eea}{\end{eqnarray}}
\def\mbf#1{\mbox{\boldmath ${#1}$}}

\newcommand{\hmol}{H$_2~$}
\begin{document}

\title{Molecular Gas Disk Structures around AGNs}

\author{Keiichi \textsc{Wada}}%
\affil{National Astronomical Observatory of Japan, Mitaka, Tokyo
181-8588, Japan}
\email{wada@cfca.jp}
\altaffiltext{1}{Department of Astronomical Science, The Graduate University for Advanced Studies,
 Osawa 2-21-1, Mitaka, Tokyo 181-8588, Japan.}
\author{Padeli P. \textsc{Papadopoulos}}
\affil{Argelander-Instit\"ut f\"ur Astronomie, Auf dem H\"ugel 71, D-53121, Germany}
\email{padeli@astro.uni-bonn.de}
\and
\author{Marco \textsc{Spaans}}
\affil{Kapteyn Institute, The Netherland}
\email{spaans@astro.rug.nl}

%




\begin{abstract}
We present new high resolution numerical simulations of the interstellar
medium (ISM) in a central $R \leq 32$ parsecs region around a
supermassive black hole ($1.3 \times 10^7 M_\odot$) at a galactic
center. Three-dimensional hydrodynamic modeling of the ISM (Wada \&
Norman 2002) with the nuclear starburst {now includes tracking of the
formation of molecular hydrogen (\hmol)} out of the neutral hydrogen
phase as a function of the evolving ambient ISM conditions with a finer
spatial resolution (0.125 pc).  In a quasi equilibrium state, mass
fraction of \hmol is about 0.4 (total \hmol mass is $\simeq 1.5\times 10^6
M_\odot$) of the total gas mass for the uniform far UV (FUV) with $G_0 =
10$ in Habing unit.  As shown in the previous model, the gas forms an
inhomogeneous disk, whose scale-height becomes larger in the outer
region.  \hmol forms a thin nuclear disk in the inner $\simeq 5~{\rm
pc}$, which is surrounded by molecular clouds swelled up toward $h
\lesssim 10~{\rm pc}$.  The velocity field of the disk is highly turbulent
in the torus region, whose velocity dispersion is $\simeq 20~{\rm
km}~{\rm s}^{-1}$ on average.  Average supernova rate (SNR) of $\simeq 5\times
10^{-5} {\rm yr}^{-1}$ is large enough to energize these
structures. {Gas column densities}  toward the nucleus larger than
$10^{22}~{\rm cm}^{-2}$ {are} observed if the viewing angle is smaller than
$\theta_v \simeq 50^\circ$ from the edge-on.  However, the column
densities are distributed {over} almost two orders of magnitude around the
average for any given viewing angle due to the clumpy nature of the
torus. For a stronger FUV ($G_0 =100$), the total \hmol mass in an
equillibrium is {only slightly smaller} ($\simeq $ 0.35), {a testimony to
the strong self-shielding nature of H$_2$}, and the molecular gas is {somewhat}
more concentrated in a mid-plane. {Other properties of the ISM
are not very sensitive either to the FUV intensity and the supernova
rate. Finally the morphology and kinematics of the circumnuclear molecular gas disks 
emerging from our models is similar to that revealed by recent near infrared
 observations using VLTI/Keck.}



\end{abstract}


\section{INTRODUCTION}
Optically thick molecular materials have been postulated to unify
various observational properties of active galactic nuclei (AGNs),
especially the two major categories of the AGN, namely type-1 and type-2
\citep{antonucci93}.  It is often assumed that the obscuration is caused
by dusty molecular gas around a supermassive black hole (SMBH), whose geometry
is schematically described as a doughnut-like `torus'
\citep[e.g.][]{urry95}.  Its spatial extent and structures are deduced
from comparison between observations in near/far infrared and
theoretical spectral energy distributions (SED) based on radiative
transfer calculations \citep[e.g.][]{fritz06,elitzur08}.  Detailed
three-dimensional radiative transfer calculations suggest that the torus
should not be uniform but clumpy \citep[][]{schart08}.  In these model
calculations, structures of the torus and its internal clumps are freely
assumed, and {then} are tested in comparison with the observed SEDs. 
For example, the inner and outer edges of the torus
are assumed typically as a sub-parsec and several tens of parcecs, respectively, and
several {hundred  dense} internal clumps are also assumed.  It was then
suggested that SEDs with the silicate absorption feature at 10 $\mu m$
are mostly affected by the inner pc region of the torus
\citep{schart08}.

Recently the dusty gas in the circum nuclear region are directly
observed in some nearby Seyfert galaxies using interferometric
mid-infrared observations.  The prototypical type-2 Seyfert NGC 1068 harbors
a warm (320 K) dust in a structure 3.4 pc in diameter and 2.1 pc in
thickness \citep{jaffe04}.  
VLTI/MIDI survey of nearby AGNs with various types recently revealed
that in the most objects AGN-heated dust is spatially compact ($\sim$ a few
pc)\citep{poncelet06,tristram07,meisen07,tristram09}.
These observations {showed} a structure of the hot
dust and gas in a central few pc region, {which} may not necessarily encompass
the whole structure of the molecular gas in the central region.
{Colder molecular gas probably extends further out, in $R\sim $ 10 pc to sub-kpc
regions. Indeed} using $^{12}{\rm CO}~(2-1)$ line observations, \citet{hsieh08} 
revealed that the molecular gas in Seyfert 1 galaxy, NGC 1097, is concentrated in $R \sim
$175 pc, {  with an} average density of the molecular hydrogen  
{  of} $n_{\rm H_2} \sim 3\times 10^4~{\rm cm}^{-3}$. Central
concentrations of high density gas around AGNs have been found via the
observations of molecular lines with high critical densities such as
${\rm HCN}~J=1-0$ $(n_{\rm crit}~=~2.3\times 10^5~{\rm cm}^{-3}$)
\citep[e.g.][]{kohno03}.

If {such quantities of} circumnuclear molecular gas {are} present, {then} star formation
 is naturally expected. In fact, there are many lines of evidence of circum nuclear
starbursts in various types of AGNs and quasars
\citep{cidfernandes04, imanishi04,davies07, maiolino07,riffel07, watabe08,chen09}.
\citet{hicks09} recently revealed structures of the ISM in local AGNs
traced by $S(1) \nu = 1-0$ line of molecular hydrogen at 2.1$\mu {\rm m}$ using
VLT/SINFONI. They found that the ISM whose scale is $r\sim 30$ pc forms
a rotating disk with a high velocity dispersion, {and} correlated
with the star formation rate.  The average gas mass is estimated to
$\sim 10^7 M_\odot$, {with a} column density of $N_{\rm H} > 2\times
10^{22}~{\rm cm}^{-2}$. They suggest that star formation {occurs within a}
 clumpy molecular gas phase. 
It has been observationally suggested that the starburst phenomena is linked with
the nuclear activities \citep{watabe08, chen09}, and 
this connection was analytically studied based on a clumpy, turbulent
 accretion disk \citep{vollmer03,vollmer04,kawakatu08,vollmer08, collin08}.

In order to construct physical models of the obscuring material around a
SMBH, {influenced also by concomitant star formation}, three-dimensional
radiative magneto-hydrodynamic calculations with a sufficiently high
resolution {and a fairly large dynamic} range (e.g.  from 100
Schwarzschild radii to 100 pc) are ultimately necessary.  However, until
now, there are a few hydrodynamic studies in the central tens pc
\citep{wada02,wada05,yamada07,schart09}.  \citet{wada02} (hereafter
WN02) investigated evolution of the ISM in the inner 100 pc region
around a $10^8~M_\odot$ SMBH using three-dimensional Euler-grid
hydrodynamic simulations, taking into account self-gravity of the gas,
radiative cooling and heating due to supernovae. They found that a
clumpy torus-like structure is naturally reproduced on a scale of tens
pc around a SMBH. The internal density, temperature are highly
inhomogeneous, and the velocity field is turbulent, but the global
geometry, i.e.  a flared thick disk, is quasi-stable, implying that
energy input from supernovae is balanced with turbulent dissipation.
They suggested that AGNs could be obscured by the surrounding
interstellar material.  These results support observational findings
that some AGNs are obscured by nuclear starbursts \citep[][and
references therein]{lev01,lev07,ballant08}.  However, the models in WN02
are not necessarily applicable to all circumnuclear regions with active
star formation, because 1) the assumed supernova rate is extremely high
($\sim 10^{-4}~{\rm yr}^{-1}~{\rm pc}^{-2}$), which would be appropriate
only for intense starburst, like ultra-luminous infrared galaxies, but
not to standard circum nuclear starbursts \citep[e.g.][]{hicks09}, and
2) the cold, dense gas in their simulations does not necessary represent
the dusty {molecular gas phase around an AGN}, because \hmol formation
and its radiative destruction by far ultra-violet radiation (FUV) from
massive stars and collisional destruction are not {included}. As a
result, the radiative cooling in WN02 is not consistent with the
chemical abundances in the low temperature medium.


Including the molecular gas in numerical models will eventually yield a
much richer interface with current and future observations. {This is because unlike
the dust continuum, the  effectively optically thin} molecular line emission
({even large line-center optical depths allow full view of turbulent media}), 
is a direct probe of the velocity fields and thus of the gaseous disk dynamics, 
unobscured by the concomitant dust. The upcoming commissioning of ALMA will drastically
enlarge such an interface by enabling the imaging of a whole suite of
molecular lines probing different density and temperature regimes at
high angular resolution.
However from the results of WN02 it is still unclear how {the
various density regimes of the H$_2$ gas}  would be distributed in the
 circumnuclear region, and what would influence its structures.  We {hereby} 
extend the three-dimensional hydrodynamic simulations of WN02 to explicitly track
 {the evolution of the \hmol phase} in the central $64^2\times32$ pc region, 
{and its interplay with the HI phase}, with a 0.125 pc resolution (cf. 0.25 pc in WN02). 
 We also {vary} the strength of the uniform FUV field and the supernova rate (SNR) {in order to
examine} their effect on the structures of molecular gas.




\section{NUMERICAL METHOD AND MODELS}
\subsection{Numerical Methods}
\begin{figure}[h]
\centering
\includegraphics[width = 8cm]{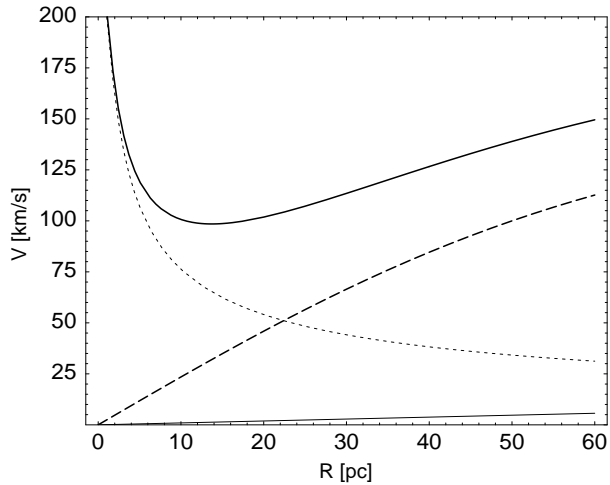} 
\caption{Assumed rotation curve (thick solid line) and contribution from each
 component of the external potential. 
Thick dashed line ($a_1 = 100~{\rm pc}, v_1 =147~{\rm km}~{\rm s}^{-1}$) and
thin line ($a_2 = 2500~{\rm pc}, v_2 =147~{\rm km}~{\rm s}^{-1}$) 
are from $\Phi_{\rm ext}$, and thick dotted line is from $\Phi_{\rm BH}$ with 
$M_{\rm BH} = 1.3\times 10^7 M_\odot$.}
\label{wada_fig: f0}
\end{figure}

We use a numerical scheme based on Eulerian hydrodynamics with a uniform
grid, which is the same as those described in
\citet{wada_norman01,wada01}.  Here we briefly summarize them.
\setcounter{footnote}{0}
We solve the following equations numerically to
simulate the three-dimensional evolution of a rotating gas disk in a
fixed spherical gravitational potential.
  \begin{eqnarray}
{\partial \rho}/{\partial t} + \nabla \cdot (\rho \mbf{v}) &=& 0,
\label{eqn: rho} \\ 
{\partial \mbf{v}}/{\partial t} + (\mbf{v}
\cdot \nabla)\mbf{v}+{\nabla p}/{\rho}  &=&  - \nabla \Phi_{\rm ext} -
\nabla \Phi_{\rm BH} - \nabla \Phi_{\rm sg}, \label{eqn: rhov}\\
 {\partial E}/{\partial t} + \nabla \cdot 
[(\rho E+p)\mbf{v}]/\rho &=& 
\Gamma_{\rm UV}(G_0) + \Gamma_{\rm SN}  -  \rho \Lambda(T_g, f_{\rm H_2}, G_0), \label{eqn: en}\\ \nabla^2
\Phi_{\rm sg} &=& 4 \pi G \rho, \label{eqn: poi} 
\end{eqnarray}
 where, 
the specific total energy $E \equiv |\mbf{v}|^2/2+ p/(\gamma -1)\rho$,
with $\gamma= 5/3$.  We assume a time-independent external potential
$\Phi_{\rm ext} \equiv -(27/4)^{1/2}[v_1^2/(r^2+
a_1^2)^{1/2}+v_2^2/(r^2+ a_2^2)^{1/2}]$, where $a_1 = 100$ pc, $a_2 =
2.5$ kpc, $v_1 = 147$ km s$^{-1}$, $v_2 = 147$ km s$^{-1}$, and
$\Phi_{\rm BH} \equiv -GM_{\rm BH}/(r^2 + b^2)^{1/2}$, where $M_{\rm
BH}=1.3\times 10^7 M_\odot$ and $b=1$ pc.  See Fig. \ref{wada_fig: f0}
for the rotation curve based on the external potentials $\Phi_{\rm ext}$
and $\Phi_{\rm BH}$.  Observationally it is hard to determine exact
rotation curves in the central 100 pc region of external galaxies, the
adopted mass distribution with a core radius of 100 pc is roughly
consistent with the rotation curves around nearby Seyfert nuclei
observed by VLTI/Keck \citep{hicks09}. CO(1-0) observations of the
central region of nearby spiral galaxies also show that their rotation
curves are steep in the central part, suggesting central massive
components \citep{sofue1999}. This could correspond to the stellar cores
found by HST/NICMOS in nearby galaxies \citep{carollo02}.  The potential
caused by the BH is smoothed inside $r \sim 1$ pc by introducing a
`softening' parameter $b$, in order to avoid too small time steps around
the BH.  This softening does not change the rotation curve in the range
shown in Fig. \ref{wada_fig: f0}, but we should not rely on the gas
dynamics and structures at $r < 1$ pc.  In the central grid cells at $r
< 0.25$ pc, physical quantities stay constant to represent a sink cell.


The hydrodynamic part of the basic equations is solved by AUSM
 (Advection Upstream Splitting Method)\citep{liou03, wada_norman01}. 
We use $512^2 \times 256$ grid points in high resolution models and
$256^2 \times 128$ grid points in low resolution models to investigate
long term evolution with different parameters\footnote{In the high
resolution models, a calculation for 4.5 Myr requires about 180 CPU
hours using 16 CPUs of NEC SX-9 whose peak performance is about 1.6
TFlops.}.  Cartesian grid points cover a $64^2 \times 32$ pc$^3$ region
around the galactic center (i.e. the spatial resolution is 0.125 pc in
the high-resolution runs and 0.25 pc for the low resolution runs). The
Poission equation, eqn. (\ref{eqn: poi}) is solved to calculate the
self-gravity of the gas using the Fast Fourier Transform and the
convolution method, where $1024^2 \times 512$ grid points and a periodic
Green's function is used to calculate the potential of an isolated
system \citep{hockney81}.

An essential progress in the present calculations compared to WN02 is
that we now solve non-equilibrium chemistry of hydrogen molecules
{along} with the hydrodynamics. Formation of \hmol on dust and its radiative and
collisional destruction are explicitly tracked, therefore
we can {deduce the}  distribution of \hmol in the central tens pc region.
See details in Appendix A.

In the energy equation (eq. (3)), we use a cooling function based on a
radiative transfer model of photo-dissociation regions
\citep{meijerink05}, $\Lambda(T_g, f_{\rm H_2}, G_0)$ $(20~{\rm K} \leq
T_g \leq 10^8~{\rm K})$ assuming solar metallicity, which is a function
of the molecular gas fraction $f_{\rm H2}$ and intensity of FUV, $G_0$
(see Appendix B for the detail). The radiative cooling rate below
$\simeq 10^4~{\rm K}$ is self-consistently {modified} depending on
$f_{\rm H_2}$, which is determined in each grid cell. We adopt heating
due to { the} photoelectric effect, $\Gamma_{\rm UV}$ and energy
feedback from type-II SNe (see below), $\Gamma_{\rm SN}$. We assume a
uniform UV radiation field, whose strength is an important parameter in
the model, represented by the Habing unit ($G_0$), the incident FUV
field normalized to the local interstellar value.

\subsection{Initial conditions and Model parameters}
The initial condition is an axisymmetric and rotationally supported thin
disk with a uniform density profile (thickness is 2.5 pc) and a total
gas mass of $M_g = 6\times 10^6 M_\odot$.  Since we allow outflows from
the computational boundaries, the total gas mass decreases during the
evolution, and it settles to $\simeq 5\times 10^6 M_\odot$ at a
quasi-equilibrium state.
Random density fluctuations, which are 
less than 1 \% of the unperturbed values, 
are added to the initial gas disk.
A uniform disk with $\rho = 1.22\times 10^3 M_\odot $ pc$^{-3}$ at $t=0$
is evolved for $t= 1.4$ Myr (the rotational period at
$R=10~{\rm pc}$ is about 0.2 Myr)
using $256^2 \times 128$ grid points (i.e. spatial resolution is 0.25 pc),
then it continues to 
higher resolution runs with $512^2 \times 256$ grid points. 
The system settles into a quasi-equilibrium state, where
the total molecular gas mass is nearly constant, after $t\sim 3.5$ Myr
(see \S 3.2 and Fig. \ref{wada_fig: f5}).

Since the current integration time is short ($\simeq 5$ Myr) due to a
limitation of our computational resources, instead of tracking the whole
life of massive stars, here we assume that supernova (SN) explosions
occur at random positions with a constant rate in the region {confined by} $r
\le 26$ pc and $|z| \le 2$ pc. {Here we effectively assumed} that massive, cold
molecular gases from which stars formed are not distributed spherically,
but they are rather concentrated in a disk with a small scale height.
Therefore we expect that supernova explosions mostly occur near the 
disk plane. 

The average SN rate (SNR) is one of {the} free parameters in the present
study.  Observationally there is a large ambiguity in the star forming
activities around AGNs. the star formation rate (SFR) around AGNs was
suggested to several tens $M_\odot {\rm yr}^{-1}~{\rm kpc}^{-2}$
\citep{davies07}.  For the Salpeter initial mass function (IMF) and
stellar mass rage of 0.1-125 $M_\odot$, type-II supernova rate is
$\simeq 0.007$ SFR, {and a} corresponding supernova rate per unit area
would be $\sim 10^{-7}~{\rm yr}^{-1}~{\rm pc}^{-2}$. For the
circumnuclear disk of $r=26$ pc in the present model, this rate suggest
that $\simeq 3\times 10^{-4} {\rm yr}^{-1}$.  On the other hand, using
the scaling relation on the star formation rate in galactic disks and in
starbursts \citep{ken98}, the initial gas density in the present model
{yields} $4\times 10^{-4}~{\rm yr}^{-1}$.  We here change the supernova
rate from 5.4 to 540 $\times 10^{-5}~{\rm yr}$.  For each type-II
supernova explosion, the energy of $10^{51}$ ergs as thermal energy and
gas mass 8 $M_\odot$ are instantaneously injected into a single cell.
Note that since we set a maximum temperature of $10^8$ K,
not all the injected thermal energy is used to
create blast waves.
Three dimensional evolution of blast
waves caused by SNe in an inhomogeneous and non-stationary medium with
global rotation are followed explicitly, taking into account the
radiative cooling, {which makes the} evolution of the supernova remnants
{depending on the} local gas density distribution around the SNe.

 We here assume a spatially uniform FUV field with $G_0 = 10$ and 100.
 For a photo-dissociation region (PDR) surrounding an HII region, $G_0
 \simeq 2\times 10^2 (L/10^4L_\odot)(r/1~{\rm pc})^{-2}$
 \citep[e.g.][]{tielens05}. For the star formation rate $10^{-5} M_\odot
 {\rm yr}^{-1}~{\rm pc}^{-2}$, which is a typical rate in starburst
 galaxies, {the} luminosity of massive stars formed for $10^7$ yr that
 contribute to the FUV field is roughly $10^4~L_\odot {\rm pc}^{-2}$. If
 we assume that the massive stars are embedded in the uniform gas disk
 with $n \sim 10^4~{\rm cm}^{-3}$, contribution from local ($\lesssim$
 a few pc) stars is important.  Then the average local FUV field in a
 grid cell of 0.1 pc would be $G_0 \sim 200$. On the other hand,
 an AGN would have a negligible effect on the FUV field in the most
 regions, because the average column density through the initial thin disk is
 $\gtrsim 10^{23}~{\rm cm^{-2}}$ for $r \gtrsim 2~{\rm pc}$. However, as
 we will see in \S 3, the star forming disk in a quasi-equilibrium state
 is highly inhomogeneous, especially in the outer region ($r \gtrsim
 5~{\rm pc}$). Therefore, $G_0$ could be several $10^3$ for the low column
 density regions ($N \lesssim 10^{22}~{\rm cm}^{-2}$) at $r \sim 10$ pc from
 a typical AGN with luminosity $L = 10^{44}~{\rm erg}~{\rm s}^{-1}$.  In
 reality, both the density field and distribution of massive stars are
 not uniform, {and} $G_0$ {should} have a large dispersion with some
 radial dependence.  Effect of the non-uniform radiation field will be
 an important subject to solve by three-dimensional radiative transfer
 calculations in the future.

 Beside the FUV, the X-ray radiation from the AGN would be
 important for the thermal and chemical structures of the ISM around an
 AGN \citep{maloney96,meijerink07}. This will be discussed in \S 5.




%

Model parameters are summarized in Table 1. For simplicity, we here
assume that the FUV intensity and SNR are independent free parameters.
{In reality those quantities are tied to the ambient star formation rate
density, which in turn is regulated by the \hmol gas density since stars
always form out of the \hmol gas. It is therefore expected that FUV, SNR
and \hmol gas mass fraction evolve in a highly coupled manner, leading
perhaps towards a self-regulating, quasi-stationary state. This
important issue can only be explored if a SF-regulating \hmol gas
distribution is included in the {ISM+stars} model, and we intend to do so in a
future paper.}
\begin{table}
  \begin{center}
    \begin{tabular}{cccc}  \hline \hline
    Model & $G_0$ & $\Delta$ [pc] &  SNR [$10^{-5}$ yr$^{-1}$] \\ \hline
     H10a  & 10  & 0.125& 5.4 \\ 
     L10a  & 10 & 0.25 & 5.4 \\ 
    H100a  & 100& 0.125 & 5.4 \\ 
    L100a  & 100& 0.25 & 5.4 \\  
    L100b  & 100& 0.25 & 54.0 \\ 
    L100b*  & 100& 0.25 & 54.0 \\ 
    L100c  & 100& 0.25 & 540.0 \\ 


\hline
     \end{tabular}
    \end{center}
  \caption{Model parameters. ``H/L'' represents ``High/Low`` resolution
  ($\Delta$, size of a numerical grid cell), ``100/10'' represents
  intensity of far ultra-violet radiation in the Habing unit ($G_0$),
  and suffix 'a-c' represent the average supernova rate (SNR). L100b* is
  the same as L100b, but energy from supernovae is injected in a larger
  scale height, i.e. $|z|\leq 10~{\rm pc}$.  }\label{index}
    \end{table}

\section{RESULTS}

\subsection{Structures of  a fiducial model}
\begin{figure}[h]
\centering
\includegraphics[width = 14cm]{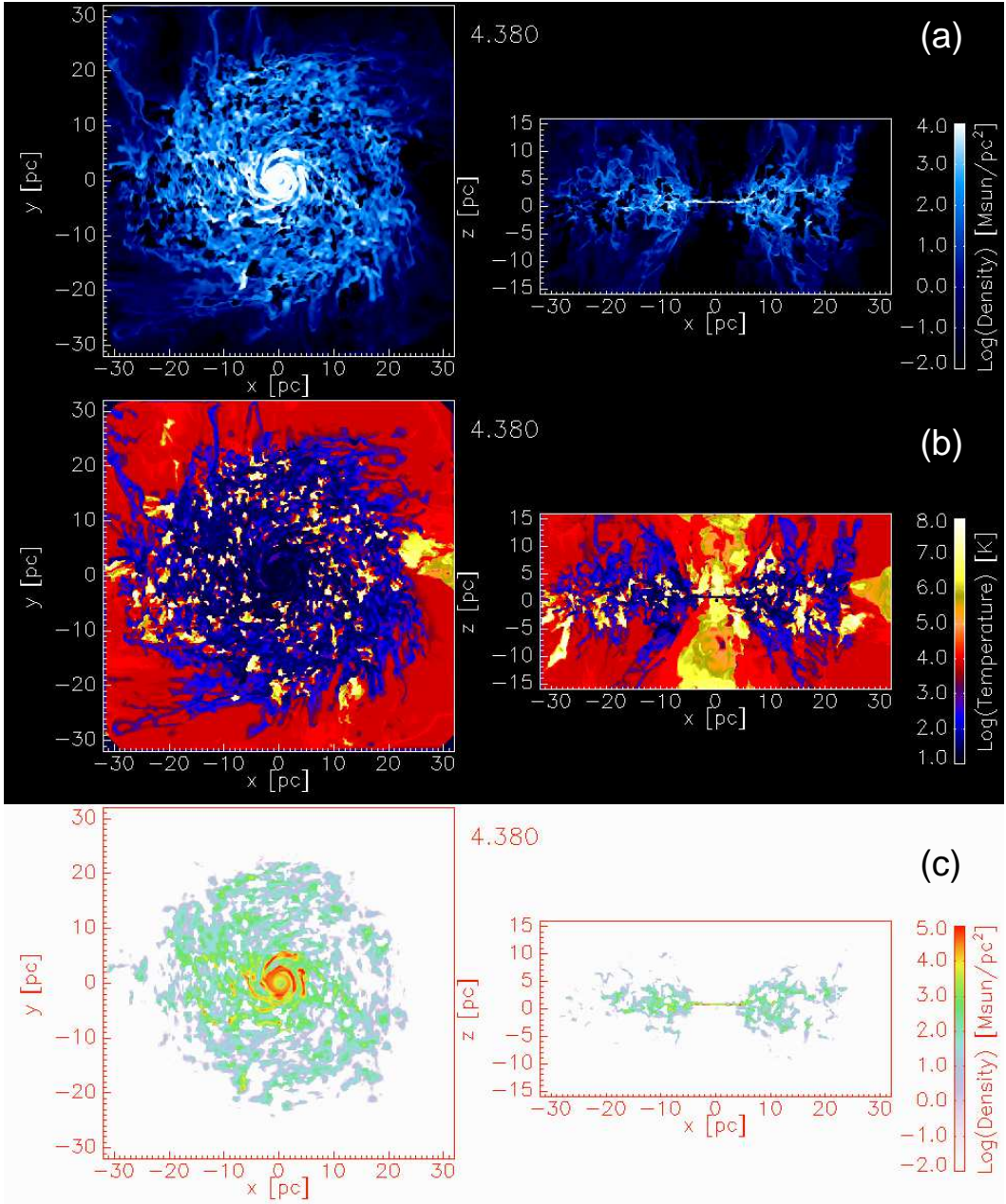} 
\caption{Cross sections  (a) gas density [$M_\odot {\rm pc}^{-3}$], (b)
 temperature [$K$] and (c) \hmol density [$M_\odot {\rm pc}^{-3}$]
on x-y and x-z planes of model H10a ($t=4.38$ Myr). }
\label{wada_fig: f1}
\end{figure}

In a quasi-steady state ($t \gtrsim 3.5~{\rm Myr}$), as reported by
WN02, the gas forms a highly inhomogeneous, flared disk, as seen in
Fig. \ref{wada_fig: f1}, which shows gas density, temperature and
molecular hydrogen density in model H100a at $t=4.38$ Myr.
 The temperature map
shows that cold ($T_g \lesssim 100$ K) gas is mainly distributed
in the high density regions.
In the central funnel-like cavity, the temperature of the gas is hot 
($T_g \gtrsim 10^6$ K). Hot gases are also patchily
distributed in the cold, flared disk. Typical size of these hot cavities is a
few pc. A large fraction of the volume is occupied by warm gas ($T_g \simeq
8000$ K). As expected distribution of \hmol roughly follows the cold, dense gas, and
therefore it forms a high density circum nuclear disk whose radius is about 5
pc, surrounded by a porous torus which extends to $\sim 5-10$ pc above
the disk plane.


\begin{figure}[h]
\centering
\includegraphics[width = 14cm]{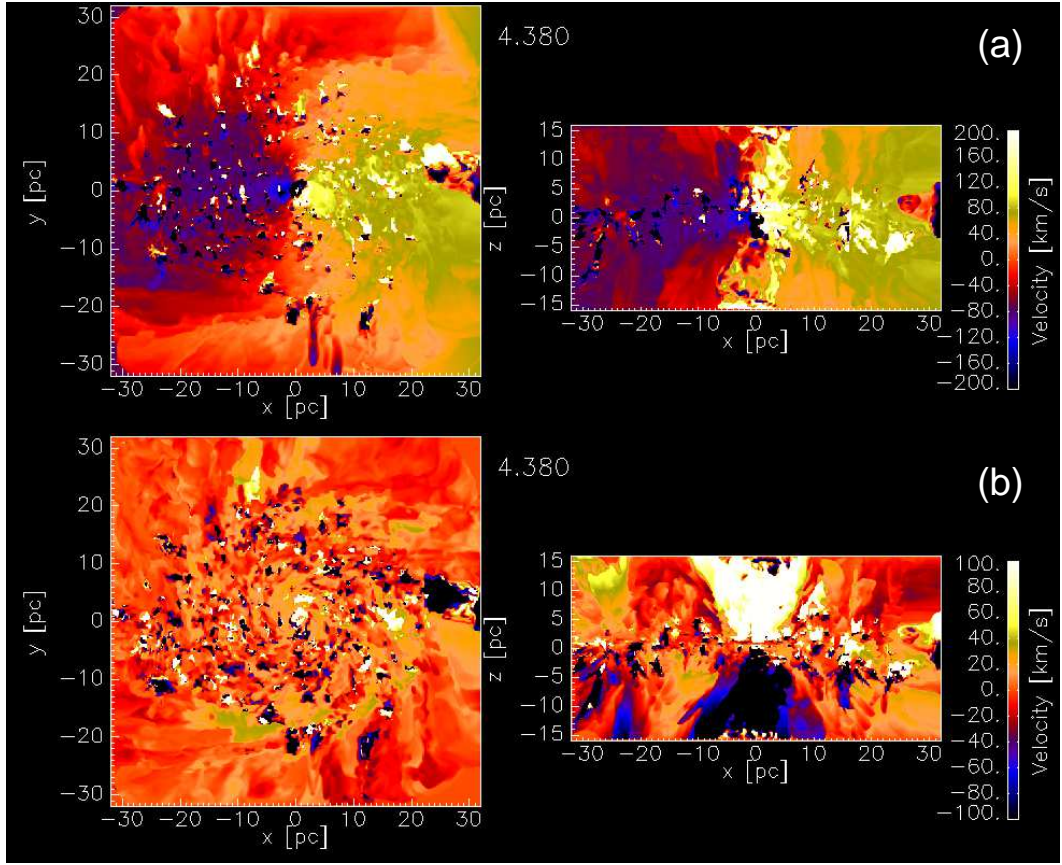} 
\caption{Same as Fig. \ref{wada_fig: f1}, but for velocity fields of the
 gas: (a) Velocity for $y-$direction, and (b) velocity for $z-$direction.}
\label{wada_fig: f2}
\end{figure}

Figure \ref{wada_fig: f2} shows velocity fields of two components ($v_y$
and $v_z$) at the same snapshot of Fig. \ref{wada_fig: f1}. The torus
shows a global rotation, but there are also large internal random
motions.  This complicated velocity field is naturally expected in an
inhomogeneous, self-gravitating disk \citep{wada_meurer02}, but the
vertical motions are mainly enhanced by energy input from supernovae
(see also Fig. \ref{wada_fig: f8}). The vertical velocity field also
shows a bipolar outflow in the central funnel, which is mostly warm and
hot gases as seen in Fig. \ref{wada_fig: f1}b.


\begin{figure}[h]
\centering
\includegraphics[width = 8cm]{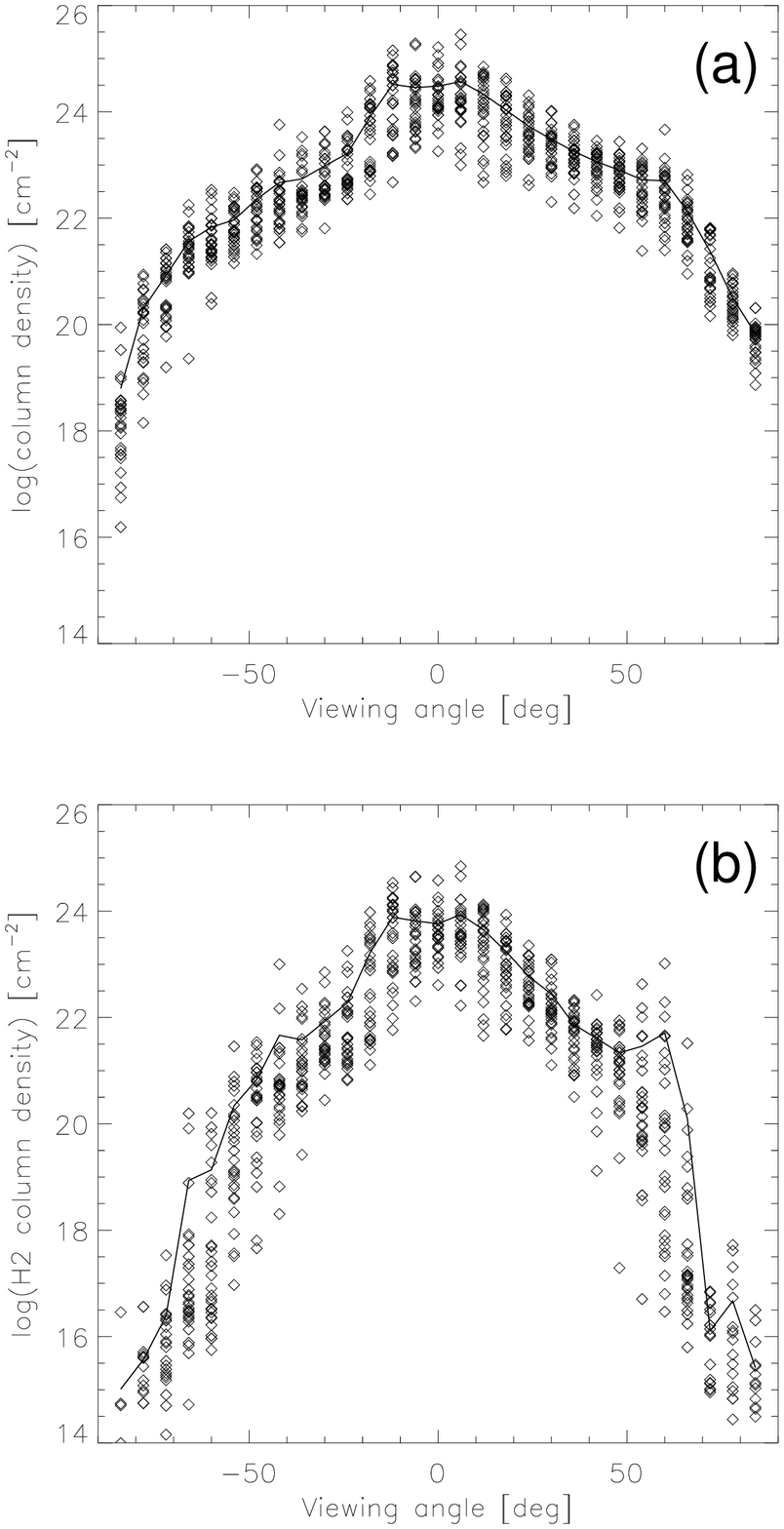}
\caption{(a) Column density of the gas, $N_{\rm g}$ and column density
 of \hmol, $N_{\rm H_2}$  as a function 
of the viewing angle in model H10a at $t=4.38 $ Myr. 
The solid line represent an azimuthally averaged column density.
}
\label{wada_fig: f4}
\end{figure}

In Figs. \ref{wada_fig: f4}a and \ref{wada_fig: f4}b, we plot total
column density ($N_{\rm g}$) and \hmol column density ($N_{\rm H_2}$)
toward the galactic center as a function of the viewing angle in model
H10a. It {clearly shows} that the total {gas} and \hmol column densities are both
largest at $\theta_v \simeq 0$ (i.e. edge-on), and the average column
density decreases {on average towards} the pole-on view {as expected}. We should note,
however that the scatter around the average is two orders of magnitude
or more for any viewing angles, which is comparable to the change of the
average value between $\theta_v = 0^\circ$ and $\pm 90^\circ$. Figure
\ref{wada_fig: f4}b shows \hmol is distributed similary, but it is more
concentrated near the disk plane.  The scatter of $N_{\rm H_2}$ is more
significant than that in $N_{\rm g}$ especially for $\theta \gtrsim
50~{\rm deg}$, reflecting that \hmol gas is highly inhomogeneous and
more sparsely distributed in larger latitudes.

\subsection{Dependence of  Model Parameters}

\begin{figure}[h]
\begin{center}
\includegraphics[width = 14.0cm]{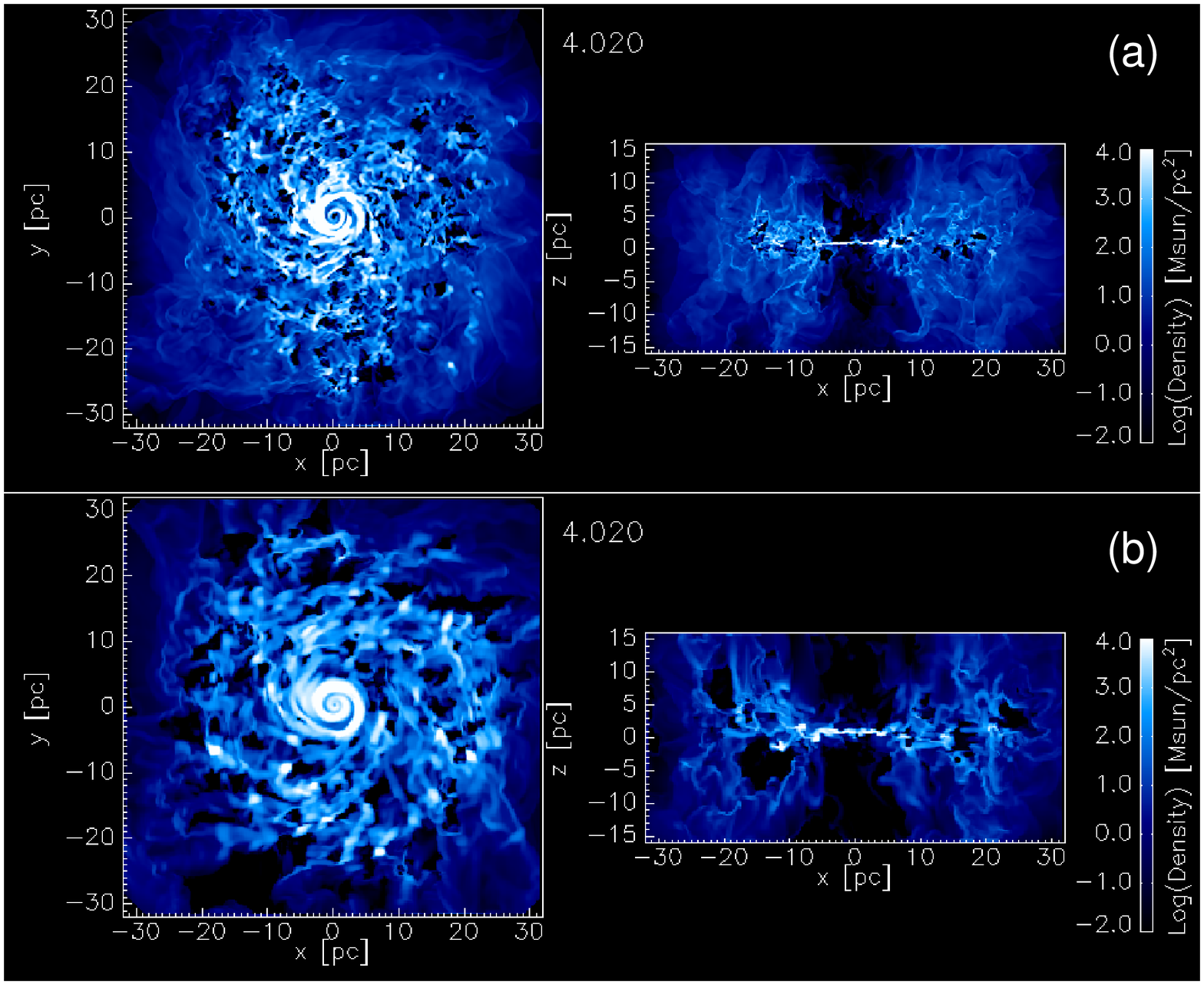}
\caption{Density distributions on x-y and x-z planes (a) in a high-resolution
 model H100a and (b) in a low resolution model L100a at $t=4.02$ Myr. } 
\end{center}
\label{wada_fig: hl100}
\end{figure}

\begin{figure}[h]
\centering
\includegraphics[width = 8.0cm]{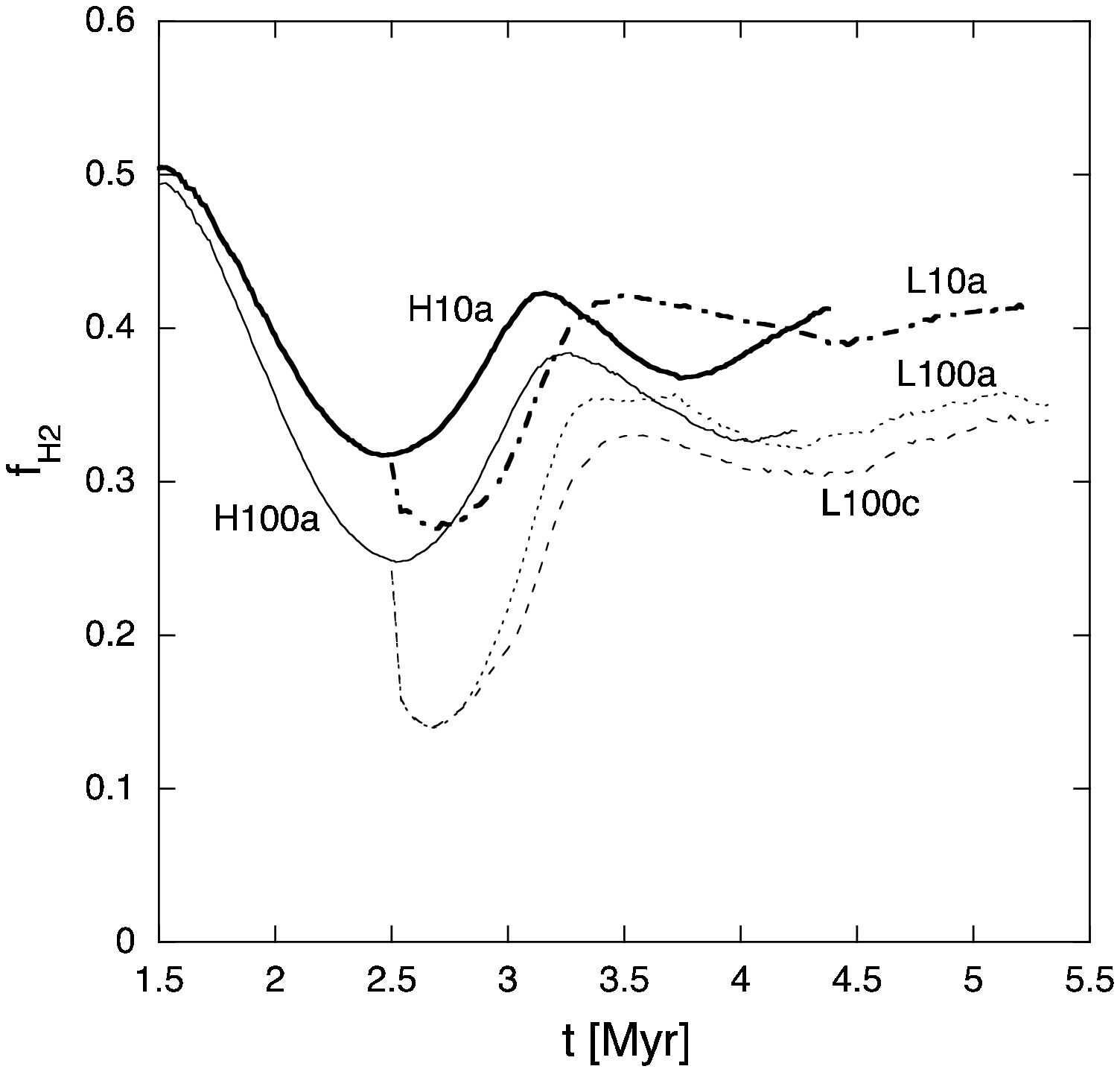} 
\caption{Evolution of molecular gas fraction $f_{\rm H2}$ in high
resolution models (H10a and H100a) and low resolution models
(L10a, L100a and L100c) recalculating from the high-resolution runs at $t=2.5$ Myr. }
\label{wada_fig: f5}
\end{figure}

Figure 5 shows density distributions of models with two resolutions,
H100a ($\Delta = 0.125$ pc) and L100a ($\Delta = 0.25$ pc) at $t=4.02$
Myr.  The inhomogeneous density fields and global morphology of the
thick disk in the two models are qualitatively similar, but clumpy
structures are finer in the high resolution model, H100a than in L100a.
In Fig. \ref{wada_fig: f5} evolution of molecular gas fractions, $f_{\rm
H_2} \equiv M_{\rm H_2}/M_{\rm total}$ in five models are shown.
The molecular gas fraction in H10a rapidly decreases from $f_{\rm
H_2}\simeq 0.5$ to $\simeq 0.32$ in the first Myr, then it increases and
slightly oscillates around $f_{\rm H_2}\simeq 0.4$.  The model with ten
times stronger FUV (H100a) shows a similar time evolution, {with
$f_{\rm H_2}$ only} about 10\% smaller than in model H10a. In order to see
a longer term evolution, we restart calculations from a snapshot of the
high resolution data at $t=2.5$ Myr with a lower resolution (i.e. 0.25
pc)\footnote{The computational time in the high resolution run is about
16 times longer than that in the low resolution runs.}. The low
resolution run (L10a) shows rapid decrease of $f_{\rm H2}$ in the first
$\sim 0.2$ Myr, {reverses to an increase} at $t\sim 2.7$ Myr, {and} then
settles around $f_{\rm H_2} \simeq 0.4$ after $t \sim 3.5$ Myr in L10a
as in H10a. Molecular fraction of L100a in the quasi-equilibrium state
is $f_{\rm H_2} \simeq 0.32$. These results suggest that for {\it a given
 supernova rate and the FUV intensity there is an equilibrium 
 \hmol fraction which does not significantly depend neither on
the numerical resolution nor on the initial state.} For stronger FUV {fields},
\hmol is less {abundant}, but this is not very sensitive to $G_0$ because
of {the effective H$_2$ self-shielding against far-UV dissociation.}
  In the low resolution runs, we change
energy input rate due to the supernova rate by a factor of 100 (see
Table 1). As seen in Fig. \ref{wada_fig: f5}, L100a and L100c show
similar evolution, but the molecular fraction is only slightly smaller in the
model with 100 times larger supernova rate.

\begin{figure}[h]
\centering
\includegraphics[width = 8.0cm]{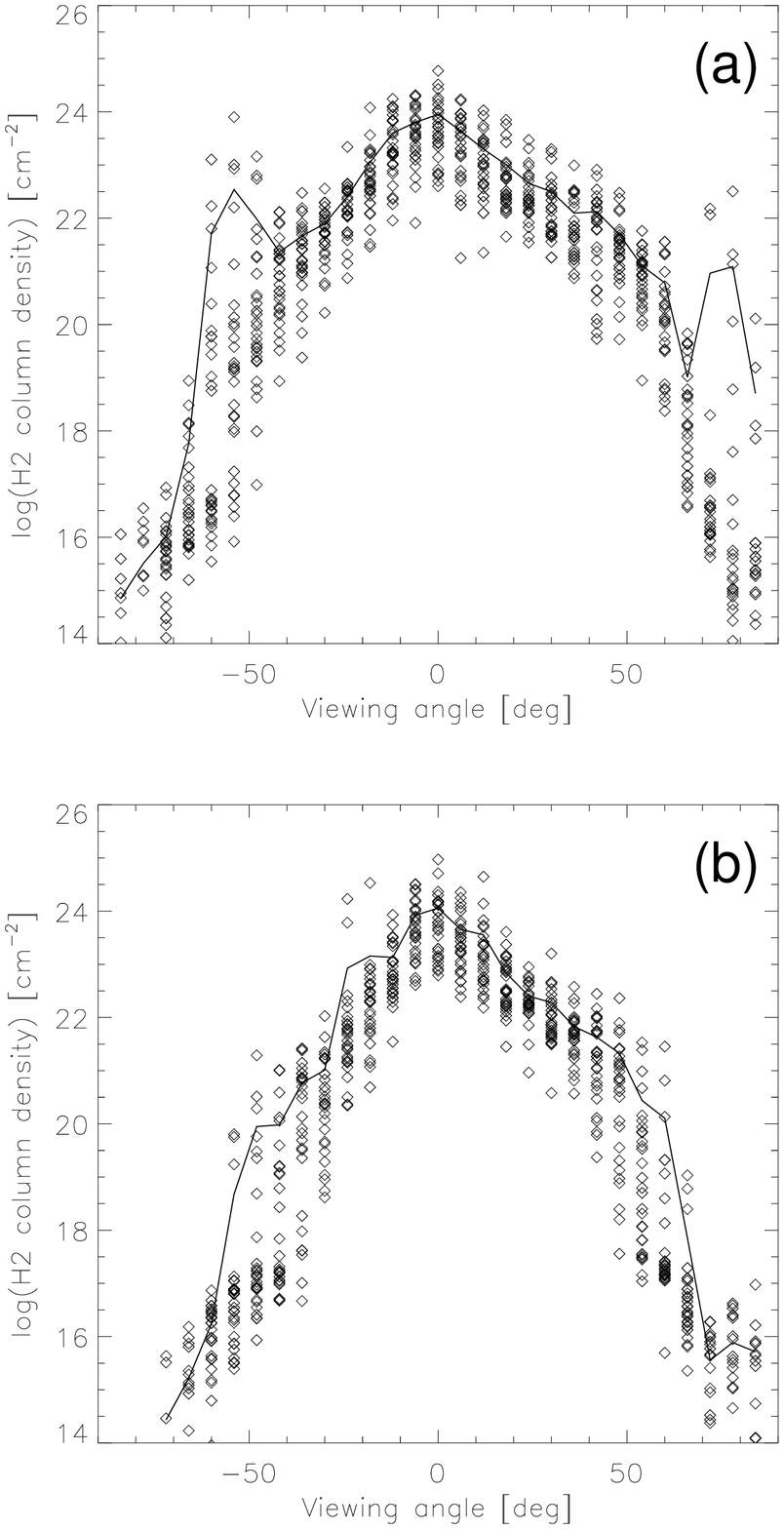} 
\caption{Column density of \hmol as a function of a viewing 
angle in H10a (Left: $G_0 = 10$) and H100a (Right: $G_0 = 100$) at $t = 3.47$ Myr. 
 The solid line represent an average column density.  }
\label{wada_fig: f7}
\end{figure}

Figure \ref{wada_fig: f7} shows column density distribution of \hmol in
models H10a and H100a at $t=3.47$ Myr. The dependence of $N_{\rm H_2}$ on
the viewing angles are similar in the two models, but in H100a, \hmol
more rapidly decreases toward high latitudes.  This is naturally
expected because \hmol is mainly formed in the dense regions, which are
concentrated near the disk plane.  Strong FUV can easily dissociate \hmol in
the relatively diffuse regions at high latitudes.  However, those
differences in the two models are not significant. Even in the strong
FUV model, \hmol is inhomogeneously distributed in the central tens pc.

\begin{figure}[h]
\centering
\includegraphics[width = 8.0cm]{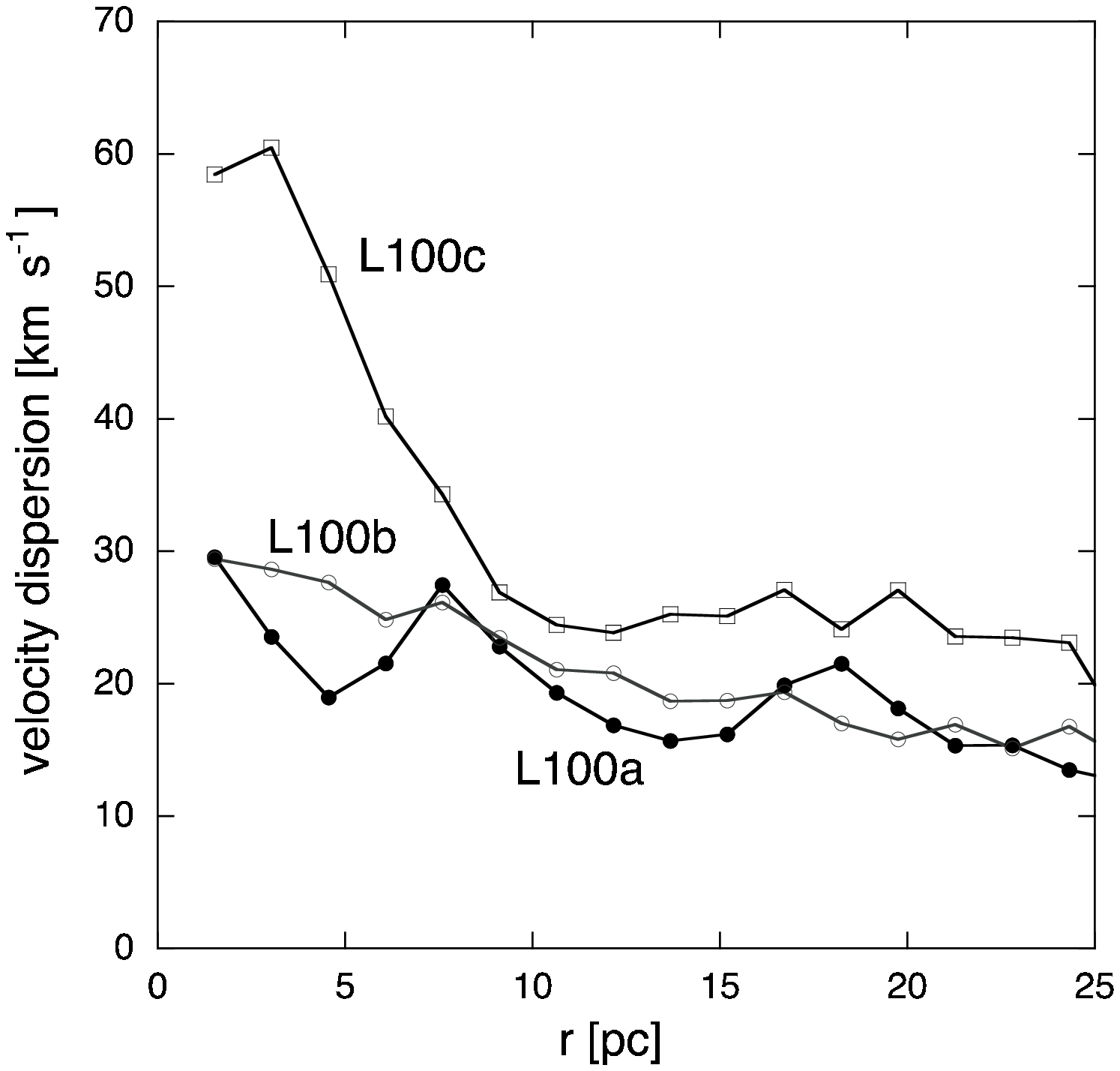} 
\caption{Azimuthally averaged velocity dispersion for $z$-direction 
as a function of radius in three models at $t=5.32$ Myr. 
Supernova rates are 5.4$\times 10^{-5}~{\rm yr}^{-1}$
 (L100a), 5.4$\times 10^{-4}~{\rm yr}^{-1}$ (L100b), and 
5.4$\times 10^{-3}~{\rm yr}^{-1}$ (L100c). }
\label{wada_fig: f8}
\end{figure}

We compare vertical velocity dispersion of the gas below $T_g = 2000$ K
in three models (L100a, L100b, and L100c) with different supernova rates in
Figure \ref{wada_fig: f8}. It shows that velocity dispersion tends to
increase with higher supernova rate. However its dependence in the torus
region is less than factor of two when the supernova rate is 100 times
larger. The effect is more significant in the inner region, where there
is a bipolar outflow from the central funnel (Fig. \ref{wada_fig: f2}).

\section{DISCUSSION}

\subsection{Comparison with Observations}

\citet{hicks09} recently revealed structures of the ISM in local AGNs
traced by a $S(1)\,\nu = 1-0$ line of molecular hydrogen at 2.1$\mu {\rm
m}$ using the near-infrared field spectrograph SINFONI at ESO Very Large
Telescope and OSIRIS at Keck Observatory. They found that {ISM with a
radius of $r\sim 30$ pc} forms a rotating disk with a high velocity
dispersion (several tens km s$^{-1}$), and suggested that star formation
is concomitant with clumpy molecular gas.  The average gas mass is
estimated to $\sim 10^7 M_\odot$, assuming a 10\% mass fraction to the
dynamical mass. The size, mass and dynamics of these nuclear molecular
gas disk around AGNs are very similar to what we found here.  The
turbulent-like velocity field revealed by their observations
qualitatively resembles to those seen in our models (Fig. \ref{wada_fig:
f2}).  They also suggest a positive correlation between the velocity
dispersion of the hot molecular gas and the star formation rate, which
is consistent with our result (Fig. \ref{wada_fig: f8}), and it is
reasonable if turbulent velocity is energized by supernovae and
{balanced by gas dissipation} (WN02).  Although $\nu = 1-0 S(1)$ is
emitted from warm gas, it is expected from our simulations that the cold
molecular gas have also similar random velocity field.

\begin{figure}[h]
\centering
\includegraphics[width = 8cm]{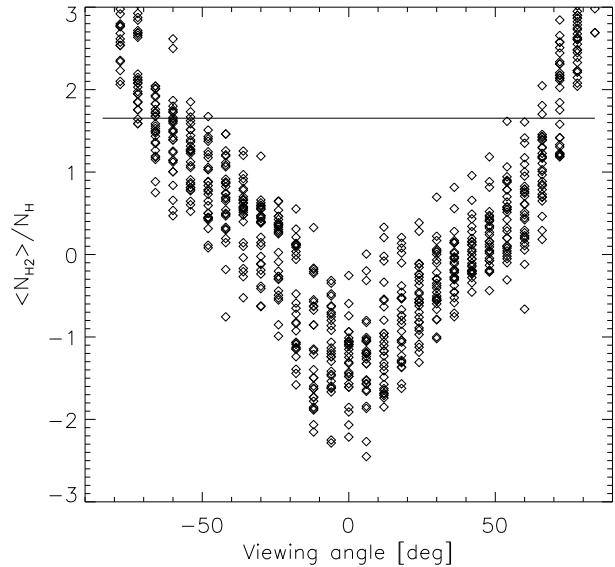}
\caption{$\langle N_{\rm H_2}\rangle/N_{\rm H}$ as a function 
of the viewing angle for model rd at $t=3.79$ Myr. The horizontal
line shows the ratio suggested by CO and X-ray observations in 
NGC 1097.}
\label{wada_fig: h2ratio}
\end{figure}

\citet{hsieh08} observed a Seyfert 1 {galaxy}, NGC 1097 { using}
$^{12}{\rm CO}(J=2-1)$ by the Submillimeter Array at angular resolution
of $4.1''\times 3.1''$ (290 pc $\times$ 220 pc), and { deduced } a total
mass of molecular hydrogen of $M_{\rm H_2} \simeq 6.5\times 10^7
M_\odot$, { with} column densities $N_{\rm H_2, CO} \simeq 5.9\times
10^{22}~{\rm cm}^{-2}$, and $N_{\rm H, CO} \simeq 1.2\times 10^{23}~{\rm
cm}^{-2}$.  They suggest a clumpy disk configuration to explain
inconsistency between the large column density suggested by their
observations and two orders of magnitude smaller column density ($N_{\rm
H, X} \simeq 1.3 \times 10^{21}$ cm$^{-2}$) suggested by X-ray
{observations} \citep{terashima02}. If the nucleus of this galaxy
surrounded by a torus consisted of high density, small gas clumps, it
would be possible to observe the broad line region through gaps between
the clumps.  On the other hand, the \hmol column density represents an
average one in the large beam size ($\sim $ 250 pc), which should be
much larger than than the value suggested by X-ray observations.  Could
this be the case in the clumpy \hmol in our results?  The ratio $N_{\rm
H_2, CO}/N_{\rm H, X}$ is $\simeq 45$ in NGC 1097. Among our models,
model H10a, in which the moderate star forming activity is assumed ($G_0
= 10$ and SNR$=5.4\times 10^{-5}$ yr$^{-1}$), would be appropriate for
comparison.
{If one supposes that the} \hmol column density inferred from the CO
line {observations} represents an average column density of all the
molecular gas present in the central several tens parsecs ($\langle
N_{\rm H_2} \rangle$), and column density of H suggested by X-ray
reflects the amount of material along the line-of-sight (its viewing
angle is $\theta_v$) for the nucleus ($N_{\rm H}(\theta_v)$).  In
Fig. \ref{wada_fig: h2ratio}, we plot $\langle N_{\rm H_2}
\rangle/N_{\rm H}$ in model H10a at $t=4.38$ Myr. 
It is clear that we have smaller chances to
observe small $N_{\rm H}$ if the line of sight is closer to the edge-on.
$\langle N_{\rm H_2} \rangle/N_{\rm H} \gtrsim 10$ can be achieved for
$\theta_v \sim 50$ deg or larger.  One should note again that the
scatter of the ratio is significant for any given viewing angles, and
it also depends on the structure of molecular gas around the nucleus.
It is therefore hard to conclude at this moment on structures of the
circumnuclear region of NGC 1097.  Future observations of AGNs using CO
or other lines with much higher resolutions, at least with a few pc
beam, by for example ALMA are necessary.

\citet{jaffe04} observed  the central region  of NGC 1068 by  the Very
Large Telescope Interferometer (VLTI), and suggested a dusty ``torus''
of thickness 2.1 pc, diameter 3.4  pc, and {temperature of} $T_g =
320~{\rm K}$.  This could correspond to the central, relatively smooth
molecular   dense   disk   typically  seen   in   our   models
(Fig. \ref{wada_fig: f1}). Due to  the strong gravity and shear caused
by the  central BH,  it is  natural that molecular  gas forms  a thin,
smooth disk  in the central few  pc. However, if  the observed central
disk  of NGC  1068 is  geometrically thick,  a strong  heating source,
e.g.  infrared radiation   would  be  necessary  \citep{pier93,  krolik07}.
\citet{tristram07} suggested a clumpy  torus plus a central thick disk
model from  {mid-IR}  observations using VLTI  of Circinus, which  has a
Seyfert 2  nucleus. Although  the size is  smaller ($\sim 2$  pc), the
geometry   and   structures  suggested   by   their  high   resolution
observations  (see a  schematic  picture  in their  Fig.  9) is  quite
similar to the models presented here.

%
%
%

\subsection{Effects of X-rays}

\begin{figure}[h]
\centering
\includegraphics[width = 8cm]{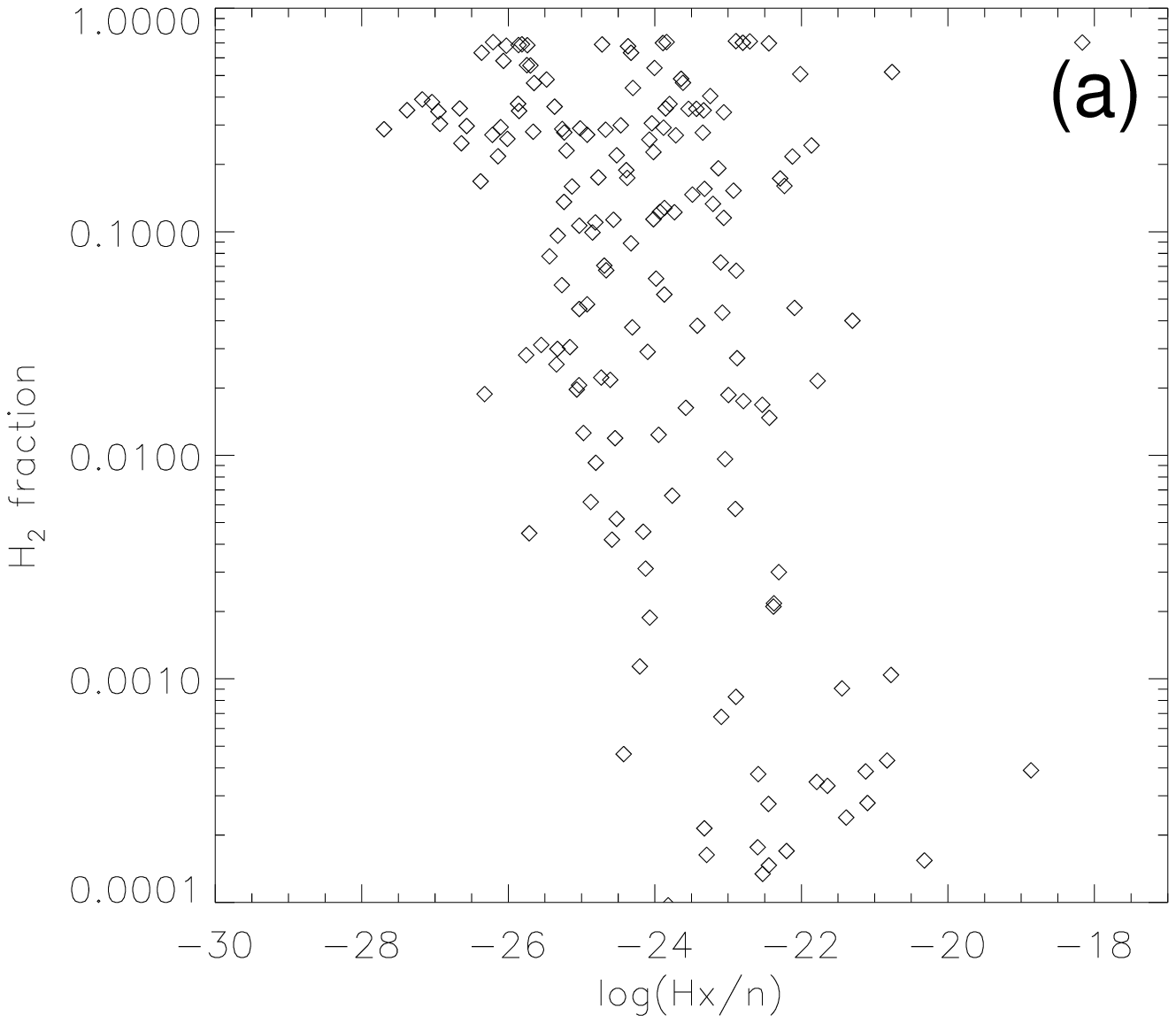}
\includegraphics[width = 8cm]{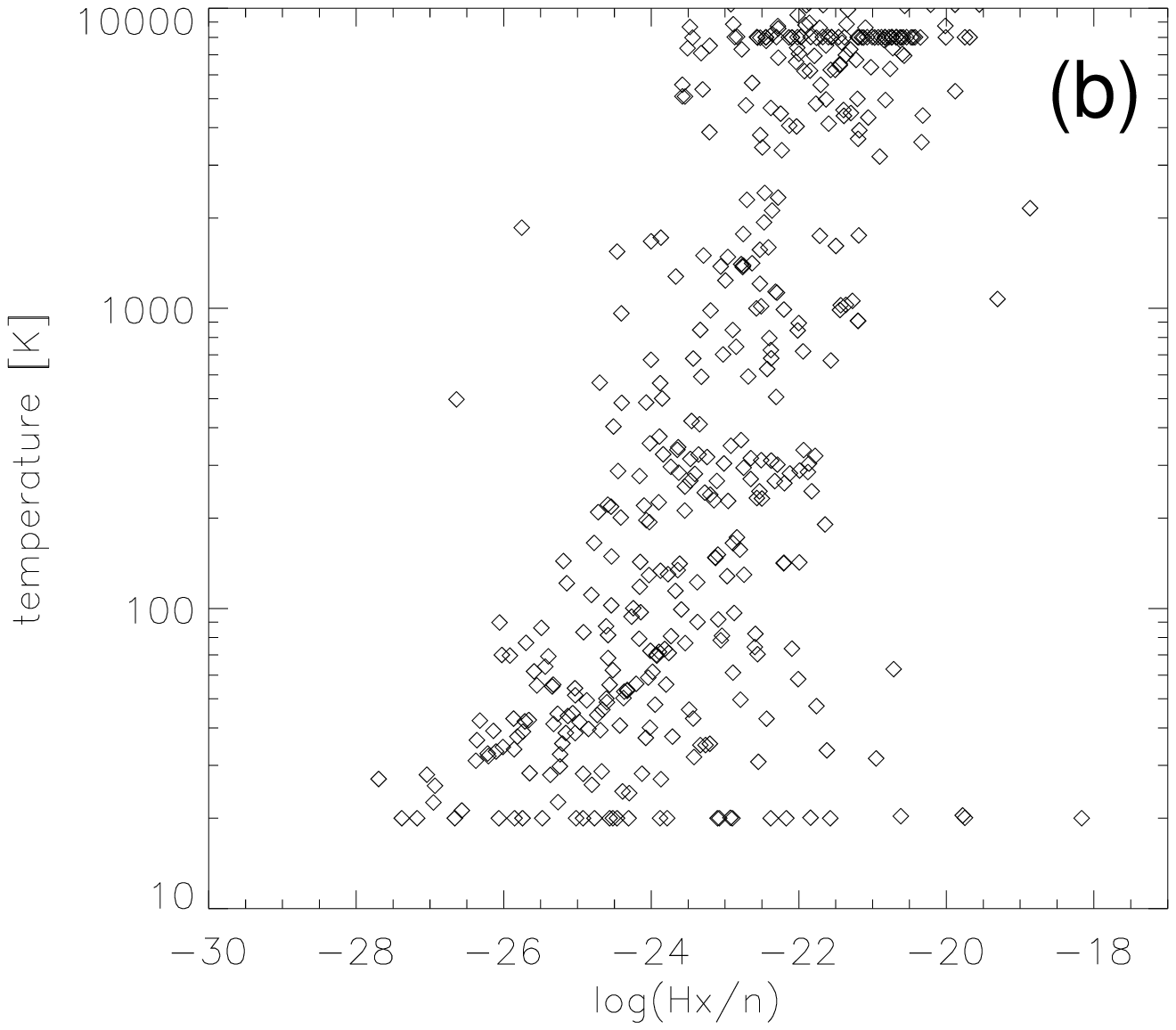}

\caption{(a) \hmol fraction as a function of ratios of the X-ray heating
deposition rate and gas density,  $H_{\rm X}/n [{\rm erg}~{\rm cm}^3~{\rm s}^{-1}]$ 
for randomly selected 500 points in the computational box in model H10a at $t=4.3$ Myr.
(b) Same as (a), but for temperature. Note that \hmol fraction is zero
 in many selected points, because the hot and warm gas occupies a large volume,
 therefore
the number of data points in the plot (a) is smaller than in (b). }
\label{wada_fig: f9}
\end{figure}

{ The strong X-ray continuum emanating from the AGN} could affect the
chemical state of the ISM in the central region
\citep{maloney96,meijerink07}.  It is known that the chemical state of
the X-ray Dissociated Region (XDR) is mainly determined by $H_{\rm
X}/n$, where $H_{\rm X}$ is a X-ray heating deposition rate and $n$ is
the number density of the gas \citep{maloney96}.  Although we do not
include the effect of X-ray from the nucleus, here we estimate its
potential effect using one of our models.  In Fig. \ref{wada_fig: f9},
we plot a fraction of \hmol and temperature as a function of $H_{\rm
X}/n~({\rm erg}~{\rm cm}^3~{\rm s}^{-1})$ at randomly selected points in
the computational box.  Here we assume that the X-ray luminosity of the
AGN is $L_{\rm X} = 10^{44}~{\rm ergs}~{\rm s}^{-1}$.  Although there is
a large scatter, for $\log(H_{\rm X}/n) \simeq -24$ or smaller the
molecule fraction increases significantly and the temperature in
most regions is less than 100\,K.  At high gas densities ($\sim
10^5~{\rm cm}^{-3}$), large ($>0.1$) \hmol fractions can be sustained
for the range of $H_{\rm X}/n$ values, limiting the impact of XDR
effects \citep{meijerink07}. This is partly because the free electrons
associated with ionization processes help to form \hmol through the
${\rm H}^-$ route and also because \hmol formation scales with density
squared.  Still, for $H_{\rm X}/n$ significantly larger than $10^{-26}$,
gas is mostly atomic as expected (Maloney et al. 1996).

These results suggest that XDR chemistry may change distribution of
\hmol around an AGN, if we explicitly include X-ray from the nucleus in
the following sense.  The \hmol abundance is indeed robust in a clumpy
medium, but the XDR gas temperature should rise relative to the PDR case
by a factor of $\sim 5$ for $\log H_{\rm X}/n > -26$
\citep{meijerink07}.  After all, X-ray ionization heating with for
example $H_{\rm X}/n \sim 10^{-25}~{\rm erg}~{\rm s}^{-1}$, is more
efficient than photo-electric emission by dust grains, therefore \hmol
gas would have higher temperatures. This leads to stronger emission in
the pure rotational \hmol lines, like {e.g. $S(0)$, $ S(1)$ at 28$\mu $m, 17$\mu $m.}
Similar considerations hold for CO \citep{spaans2008}. Shocks can heat as
well, therefore observational data on line profile kinematics would be
useful in the future.


\subsection{A self-regulated \hmol and the star formation rate}
In the present results it is notable that although the assumed FUV and
supernova rate differ by orders of magnitude among the models, total
molecular gas fraction settle between $f_{\rm H_2} \simeq
0.3-0.4$. There is a weak `negative' feedback of FUV and supernovae on
the star formation in the central region. This suggests that the
molecular gas disk could survive around AGNs associated with a
starburst.  Since the negative feedback of FUV and supernova rate on
\hmol mass is `weak', it is also implied that if there is an episodic
large mass inflow into the circum nuclear region, we expect a burst of
star formation, because star formation rate depends on \hmol density.
This could also affect the mass accretion rate toward the AGN via
turbulent viscosity \citep{kawakatu08}. {To explore these mechanisms,
the star formation, the radiative feedback from the AGN/stars affecting
\hmol (the star formation fuel), and a coupled SNR should be
self-consistently followed \citep[e.g.][]{wada08} for at least several
tens Myr. These effects, along with incorporating the \hmol-tracing CO
molecule, will be investigated in a future paper.}

\subsection{Positions of supernovae}
\begin{figure}[h]
\centering
\includegraphics[width = 10.0cm]{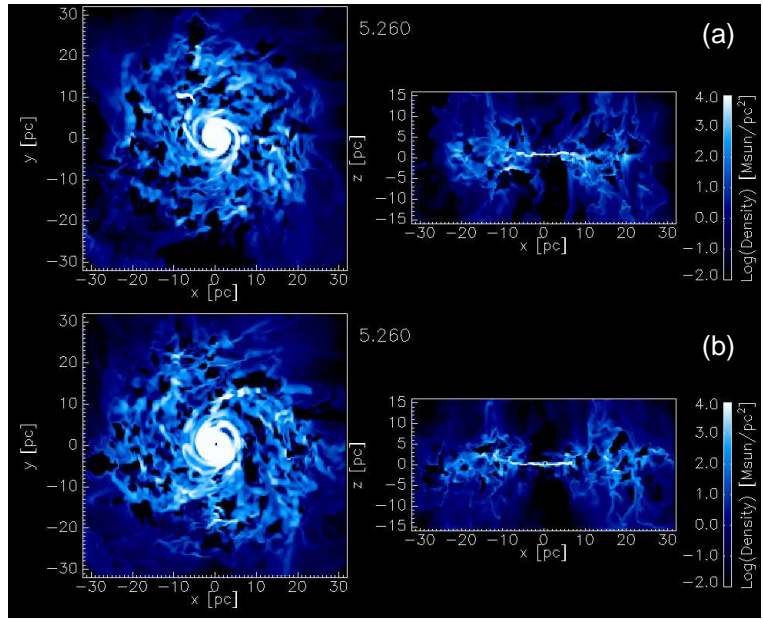} 
\caption{Density distributions in models with different positions of supernovae,
 $|z|\leq 2$ pc (L100b) and $|z|\leq 10$ pc (L100b*). }
\label{wada_fig: pos}
\end{figure}

Finally, effect of position of supernovae on the gas density
distribution is compared between L100b and L100b* (Fig. \ref{wada_fig:
pos}). As explained in \S 2.2, we assumed that SN explosions occur at
random positions with a constant rate in the region confined by $|z|
\leq 2$ pc. This is justified by the results that \hmol is more
concentrated in the disk plane. However, orbits of massive stars formed
near the disk plane could have large inclination angles, as a result,
energy from supernovae could be released at higher latitudes.  In model
L100b*, supernovae are assumed to be exploded in $|z| \leq 10$ pc
(Fig. \ref{wada_fig: pos}b), but we do not see significant differences
in the distribution of gas around AGNs.

\section{CONCLUSIONS}

We present new high resolution numerical simulations of the interstellar
medium (ISM) in a central $R \leq 32$\,pc region around a supermassive
black hole ($1.3 \times 10^7 M_\odot$) at a galactic
center. Three-dimensional hydrodynamic modeling of the ISM (Wada \&
Norman 2002) {along with a concomitant} starburst {now includes tracking
of the formation of molecular hydrogen (\hmol)} out of the neutral
hydrogen phase as a function of the evolving ambient ISM conditions with
a finer spatial resolution (0.125 pc).  Radiative cooling rate is
self-consistently changed depending on abundance of \hmol in each grid
cell.  In a quasi equilibrium state, mass fraction of \hmol is about 0.4
(total \hmol mass is $\simeq 1.5\times 10^6 M_\odot$) of the total gas
mass for the uniform far UV (FUV) with $G_0 = 10$ in Habing unit.  As
shown in the previous model, the gas forms an inhomogeneous disk, whose
scale-height becomes larger in the outer region.  \hmol forms a thin
disk in the inner $\simeq 5~{\rm pc}$, and it is clumpy and swelled up
toward $h \simeq 10~{\rm pc}$.  The velocity field of the disk is
highly turbulent in the torus region. Outflow of the gas forms a hot
($T_g \sim 10^6$ K)
funnel in the inner region.  These structures and dynamics are energized
by supernovae with average supernova rate (SNR) of $\simeq 5.4\times
10^{-5} {\rm yr}^{-1}$. { Gas column densities} toward the nucleus
larger than $10^{22}~{\rm cm}^{-2}$ {are} found if the viewing angle
is smaller than $\theta_v \simeq 50^\circ$ from the edge-on.  However,
the column densities are distributed {over} almost two orders of
magnitude around the average for any given viewing angle due to the
clumpy nature of the torus. {The column density} of \hmol shows a
similar distribution, and $N_{\rm H2} \gtrsim 10^{22}~{\rm cm}^{-2}$ is
achieved for $\theta_v \lesssim 30^\circ$.  For stronger FUV ($G_0 =
100$), the total \hmol mass in an equilibrium is {only a little} smaller
($\simeq $ 0.35) and the molecular gas is slightly more concentrated in
a plane, while other properties of the ISM {are also not very}
sensitive to the FUV intensity and supernova rate. Vertical velocity
dispersion of the gas in the torus ($\simeq 20~{\rm km}~{\rm s}^{-1}$ on
average) is enhanced {only} by a factor of $\sim$ 1.5 for increasing the
supernova rate by two orders of magnitude.

The structure and dynamics of these nuclear molecular disk presented
here are very similar to molecular gas disk in various types of nearby
AGNs recently revealed by ro-vibration line of \hmol using VLTI/SINFONI
and Keck/OSIRIS \citep{hicks09}. The inconsistency on column density
toward the center in NGC 1097 between the large column density suggested
by CO observations \citep{hsieh08} and X-ray \citep{terashima02} could
be {attributed to the inhomogeneous structures emerging from the hereby
presented disk models.}

Finally, {in the present results it is notable} that although the assumed
FUV and supernova rate may differ by orders of magnitude among the models,
the total molecular gas fraction settles between $f_{\rm H_2} \simeq
0.3-0.4$. {This weak `negative' feedback from FUV and supernovae
on the star formation in the central region, suggests that substantial
molecular gas disks can survive around AGNs with strong ongoing star formation. }


%
\vspace{0.5cm}
%
\acknowledgements

{We thank the anonymous referee for his/her valuable comments.  We are
also grateful to Nozomu Kawakatu and Masako Yamada for helpful discussions.  The numerical
computations presented in this paper were carried out on NEC SX-9 in the
Center for Computational Astrophysics, NAOJ. This work is partly
supported by a collaborating project between NAOJ and NEC}.




\setcounter{section}{0}
\renewcommand{\thesection}{A\arabic{section}}

\setcounter{equation}{0}
\renewcommand{\theequation}{A\arabic{equation}}

\setcounter{figure}{0}
\renewcommand{\thefigure}{A\arabic{figure}}

\appendix
\section{Numerical recipe of \hmol formation and destruction}

Abundance of hydrogen molecule of each hydrodynamic grid-cell (0.125 or 0.25
pc) at a given time step is determined by the following time-dependent 
approach coupled with hydrodynamic equations.
  In this approach \hmol and HI are assumed fully mixed within ``cells'' of
 typical radius $R_{\rm c}$
whose choice is motivated by the ISM physics, while
it remains unresolved by the simulation during the course of the
thermodynamic and dynamic evolution of the gas.
This allows a smooth treatment of the HI-\hmol mass exchange in both 
the Cold Neutral Medium (CNM) and Warm Neutral Medium (WNM) phases,
 with all factors now
  incorporated in a single equation, namely

\begin{eqnarray}
 \frac{dn_2(\vec r)}{dt}&=&  R_{\rm f}(T_{\rm k}) n(\vec r)^2 \nonumber \\ 
&-&\left[2R_{\rm f}(T_{\rm k}) n(\vec r)+ \frac{G_0k_{d}}{4\pi}
\int _{4\pi } f_s(N_2)e^{-\tau (\Omega )}d\Omega\right] n_2(\vec r) \nonumber  \\
&-& \gamma _1(T_{\rm k}) \left[n(\vec r)-2n_2(\vec r)\right]n_{2}(\vec r)-\gamma_2(T_{\rm k})n_2(\vec r)^2.
\end{eqnarray}
The  factor  $ k_{\rm d}= 4\times  10^{-11}\,  {\rm s}^{-1}$  is  the
unshielded FUV-induced \hmol dissociation rate, $  G_{0}$ is the
FUV  radiation  field  in  Habing  units ($1.6\times 10^{-3}~{\rm
ergs}~{\rm cm}^{-2}~{\rm s}^{-1}$), $\tau$ is the FUV optical depth,  $   n(\vec  r)=n_1(\vec
r)+2n_2(\vec r)$  ($n_1$: neutral hydrogen density, $  n_2$:  \hmol density),
and $  \gamma  _1(T_{\rm k})$, $  \gamma _2 (T_{\rm k})$  are the H-\hmol
and \hmol-\hmol  collisional destruction rates of \hmol for 
gas temperature $T_{\rm k}$  that can be
fitted using data from \citet{martin98}.
The function $R_{\rm f}(T_{\rm k})$ is the \hmol formation  factor.
We set $R_{\rm f} = 3.5\mu\times 10^{-17} (T_{\rm k}/100~{\rm
K})^{1/2} S_{\rm H}(T_{\rm k})~{\rm {\rm cm}^3~{\rm s}^{-1}}$ \citep{jura75,pelupe06}
for $T_{\rm k} \leq 500$ K, otherwise $R_{\rm f}  = 0$.
{  The factor $\mu$ represents obsevational uncertainties in the \hmol formation
rate, with the canonical value \citep{jura75} corresponding to
$\mu =1$, but values up to $\mu =5$ are possible  \citep[see][]{papado02}.}
Here we adopt $\mu = 5$, since this seems more probable form ISO studies
of
Galactic Photodissociation regions \citep[][and references
therein]{papado02}. 
 {The function $ S_{\rm H}(T_{\rm k})$ quantifies
the sticking probability of an H atom on a grain, which we set to unity
for most of the modeling done in this work (the \hmol formation probability
after sticking is considered unity throughout). We  only test the case
of $S_{\rm H}=\left(1+T_{\rm k}/T_{\circ}\right)^{-2}$ where $T_{\circ} = 100\,K$
which was obtained by \citet{BuZh91}, and is thought to be good for dust
grains in general, as long as they are covered by a few layers of weakly bonded
material.} 

{  The  $f_{\rm s}(N_2)$ function denotes the \hmol self-shielding factor, with $N_2$  
being the \hmol column density, and}

\begin{eqnarray}
  f_{\rm s} &=& 1, \;\;\;\;\;\;\;\;\;\;\;\;\;\;\;\;\;\;\;  {\rm for} \;\; N_2\leq N_0 = 10^{14}~{\rm cm}^{-2}, \nonumber \\
&=&  \left(\frac{N_{2}}{N_0}\right)^{-k}\;\;\;\;\;\; {\rm for} \;\; N_2>N_0
\end{eqnarray}
{   where $k$=3/4 (see \citet{draine96}).}

If we  integrate eq. (A1)  over the  volume $  \Delta  V_{\rm c}$ of  the cell,
while  assuming  negligible  dust  absorption  within  it  ($   \tau
(\Omega)\sim 0$) we obtain,

\begin{eqnarray}
  \frac{dn_2}{dt}&=&   R_f(T_{\rm k}) n^2-\left[2R_f(T_{\rm k}) n+
  G_0 k_{\rm d}\left(\frac{N_{0}}{n_2 R_{\rm c}}\right)^k K(R_{\rm c})\right] n_2- \nonumber \\
&&  \gamma _1(T_{\rm k}) \left(n-2n_2\right)n_{2}-\gamma_2(T_{\rm k})n_2 ^2,
\end{eqnarray}

\noindent
where the factor $  K (R_{\rm c})$ ($R_{\rm c}$ is a cloud radius) is

\begin{eqnarray}
  K(R_{\rm c}) &=&\frac{1}{4\pi\, \Delta V_{\rm c}}\int _{\Delta V_{\rm c}} \int _{4\pi} \left(\frac{R_{\rm c}}{|\vec R_{\rm c} 
-\vec r|}\right)^k d\Omega d^3 \vec r \\
  &=& \frac{3}{4-2k}\int ^{1} _{0} x\left[(1+x)^{-k+2}-(1-x)^{-k+2}\right] dx,\, \;\;\;{\rm where}\, \;\;\;x=r/R_{\rm c}, \\
  &=& \frac{3\times 2^{-k+3}}{(3-k)(4-2k)}\left[\frac{2(3-k)}{4-k}-1\right],
\end{eqnarray}
and for $ k=3/4$ results to $ K(R_{\rm c})=0.96$.  Unlike
\citet{pelupe06}, the description of the HI-\hmol gas mass
exchange here does not involve the density-size power law.  It utilizes only
the much better established line-width-size scaling law found for
molecular clouds \citep{heyer04}, and used here to define the
``cell'' size (A.7) within which HI and \hmol are assumed fully mixed.
Aside from the much smoother implementation of the \hmol formation and
destruction mechanisms that this allows, it yields a treatment that
remains suitable for a gas phase that may not follow both scaling
relations (i.e. diffuse and non self-gravitating gas).

  The assumption of a typical HI-\hmol mixing sub-grid ``cell" where
  \hmol formation and dissociation takes places in full concomitance
  with HI greatly simplifies the problem of tracking the HI$\Leftrightarrow$\hmol gas
  phase exchange by reducing a non-local radiative transfer problem to a
  local one. Thus the \hmol fractions estimated are likely to be lower
  limits since \hmol self-shielding from FUV radiation occurring between
  distant ``cells" (or dust absorption of the FUV radiation over scales
  larger than 2$R_{\rm c}$) can only result to higher \hmol fractions.

  The local approximation made here in order to simplify a formidable radiative
transfer problem in turbulent media still leaves the choice of the HI-\hmol
 mixing scale $R_{\rm c}$ open. A natural choice of $R_{\rm c}$ for our
 three dimensional hydrodynamic+thermodynamic ISM models would be 
$  R_{\rm c}\sim  1/2~ l_{p}$,  where $   l_{p}$ 
is the scale where the turbulent ISM pressure begins to dominate the
thermal pressure, or equivalently the scale where the cloud size
$R_{\rm c}$-linewidth ($\Delta V$) relation breaks down and $   \Delta
V(l_p)\sim  \Delta V_{\rm th}$. The corresponding
radius is then given by:

\begin{equation}
  R_{\rm c} = 0.5 \left( \frac{\Delta V_{\rm th}}{\Delta V _0}\right)^{1/q}~{\rm pc}, \, \;\;\;\; {\rm with}\,  \;\;\;\; \Delta V_{\rm th}\simeq 0.8~(T_{\rm k}/100~{\rm K})^{1/2}\,~ {\rm km}~{\rm s}^{-1},
\label{eq: a7}
\end{equation}

\noindent
where   $   \Delta   V(l)=\Delta   V_0  (l/{\rm pc})^q   $  is   the
size-linewidth relation found for  the cool ISM. Several observational
and theoretical studies favor $  q\sim 1/2$ \citep[e.g.][]{larson81,elmegreen89,heyer04}.
The normalization factor $  \Delta V_0=1.2~{\rm km}~{\rm s}^{-1}$ for both
CNM and WNM \citep{wolfire03}.
Below such scales the thermal velocity fields take over, setting a natural
resolution limit for any hydrodynamic ISM models tracking macroscopic
gas motions. Moreover tracking the thermal state of the gas  as a function
of time and position allows $R_{\rm c}$ to adjust accordingly (cf. eq. \ref{eq: a7}), as
expected in the true multiphase medium \citep[e.g.][]{chieze87}. 
Finally, by avoiding linking $R_{\rm c}$ to the numerical resolution
of our hydrodynamic models \citep[e.g.][]{glover07} keeps the computed
 \hmol gas mass fraction independent of it.
We set a maximum value of $R_{\rm c}$ corresponding to the scale
length of $T_k \sim 500~{\rm K}$ gas
 (cf. eq. \ref{eq: a7}). This temperature limit
broadly marks the CNM$\rightarrow$ WNM thermodynamic transition zone
beyond which purely thermal motions fully overtake the
macroscopic ones over the scales resolved here, while
little \hmol content is expected ($R_f\sim 0$ for $T_k>500~{\rm K}$ { while
 collisional destruction of  \hmol becomes very efficient}).

In  the case  of CNM-to-WNM  or the  WNM gas  phase $   R_{\rm c}$  can be
substantial, and the dust extinction term can be important. Then we use

\begin{equation}
  K(R_{\rm c})=\frac{1}{4\pi\, \Delta  V_{\rm c}}\int _{\Delta V_{\rm c}} \int _{4\pi}
\left(\frac{R_{\rm c}}{|\vec R_{\rm c} -\vec r|}\right)^k e^{-n \sigma _{\rm FUV}|\vec
  R_{\rm c}- \vec r|} d\Omega d^3 \vec r,
\end{equation}

\noindent
which after some simplifications results to,

\begin{equation}
  K(R_{\rm c})=\frac{3}{2} \int ^1 _0 x \left[\int ^{1+x} _{1-x}
t^{1-k} e^{-(n\sigma _{\rm FUV} R_{\rm c}) t} dt \right] dx,\, \;\;\;{\rm where}\, \;\;\;x=r/R_{\rm c}.
\end{equation}
The  latter  expression encompasses  both  dust  extinction and  \hmol
self-shielding and is the correct one for use in eq. (A3).  For $  \sigma
_{\rm FUV}=0$,  $ K(R_{\rm c})$ becomes identical  to that given in  eq. (A6). In
the case of pure dust  shielding (i.e. $k=0$), 
and after some calculations it
becomes

\begin{equation}
  K(R_{\rm c})=\frac{3}{2} \left[\frac{(2\alpha -3)+(2\alpha ^2 + 
4\alpha+3)e^{-2\alpha }}{\alpha ^4}\right],\,  \alpha = n\, \sigma _{\rm FUV} R_{\rm c}.
\end{equation}

\section{Cooling Function based on PDR models}

The cooling curves in temperature ($T_g$), H$_2$ abundance and radiation field
strength $G_0$ are derived from a grid of photo-dissociation region
(PDR) models that have a range in irradiation ($G_0=0-10^4$ values) and
density $1-10^6~{\rm cm}^{-3}$.  The radiation field extends beyond 13.6
eV according to a starburst99 spectrum for a Salpeter IMF, i.e., the HII
region is computed as part of the PDR.  The code of \citet{meijerink05},
with the latest chemical rates and grain surface reactions as in
\citet{cazaux09} is adopted.

The PDR models, for a given $G_0$ and density, are evaluated at the
particular H$_2$ abundances and temperatures that they enjoy with depth
to get a parameterization of the cooling rate in $G_0$, H$_2$ abundance
and $T_g$. In general, the chemical and thermal balance in PDR models
depends on the ambient density. As it turned out, a cooling curve
extraction (density squared scaling) was possible because many cooling
lines have critical densities well in excess of $10^3$ cm$^{-3}$.

However, optical depth effects can suppress cooling and drive level
populations to LTE.
Therefore, information on the typical column over which coolants exist
chemically in a PDR are used to correct for line trapping, as follows.
A local turbulent velocity dispersion (Gaussian $\sigma$) of $dV=3~{\rm
km}~{\rm s}^{-1}$ is adopted and some turbulent coherence length $L$ (between 3 and 0.3 pc).
This yields a formal velocity gradient $dV/L$ for scales larger than $L$.
If the opacity is concentrated on small physical scales, then radiation
is trapped. Still, different parts of the PDR cloud do not see each other
on large physical scales. In effect, one has an effective Sobolev
approximation or not depending on the chemical stratification. We adopt
$L=0.3$ pc in these simulations to mimic the effects of a turbulently
active nuclear region. In this, the multi-zone escape probability code
of \citet{poelman05} is used for the radiative transfer in the
cooling lines (under statistical equilibrium). Each zone in a PDR was
treated this way, and a correction on the cooling curve due to line
trapping effects can be derived.

For the warm and dense gas in the simulation clumps, the opacity is
concentrated on small physical scales, leading to line trapping.  For
example, cooling lines like [CII] 158 $\mu$m, [OI] 63 $\mu$m and CO all
have optical depths significantly larger than unity across a single PDR
interface. Given that this PDR sub-structure is not resolved in the
simulations, we feel that these opacity effect should be included.

The dust temperature is computed as part of the PDR model and is
often larger than 30 K. The far-infrared radiation field produced by
this dust is included in the statistical equilibrium of all cooling
agents. Particularly water is quite sensitive to dust pumping and is
an effective heater in the presence of dust grains warmer than 50 K.

At temperatures above $10^4$ K the presence of an ionizing component
in the impinging radiation field was found to suppress the cooling rate,
compared to a collisionally ionized plasma, by a factor of $\sim3$
for $T=10^4-10^5$ K. This is of little concern in our hydrodynamical
simulations. Line transfer in the resonantly scattered Lyman $\alpha$
line benefits very strongly from dust absorption and re-radiation in
the optically thin far-infrared continuum.

\begin{figure}[h]
\centering
\includegraphics[width = 7.5cm]{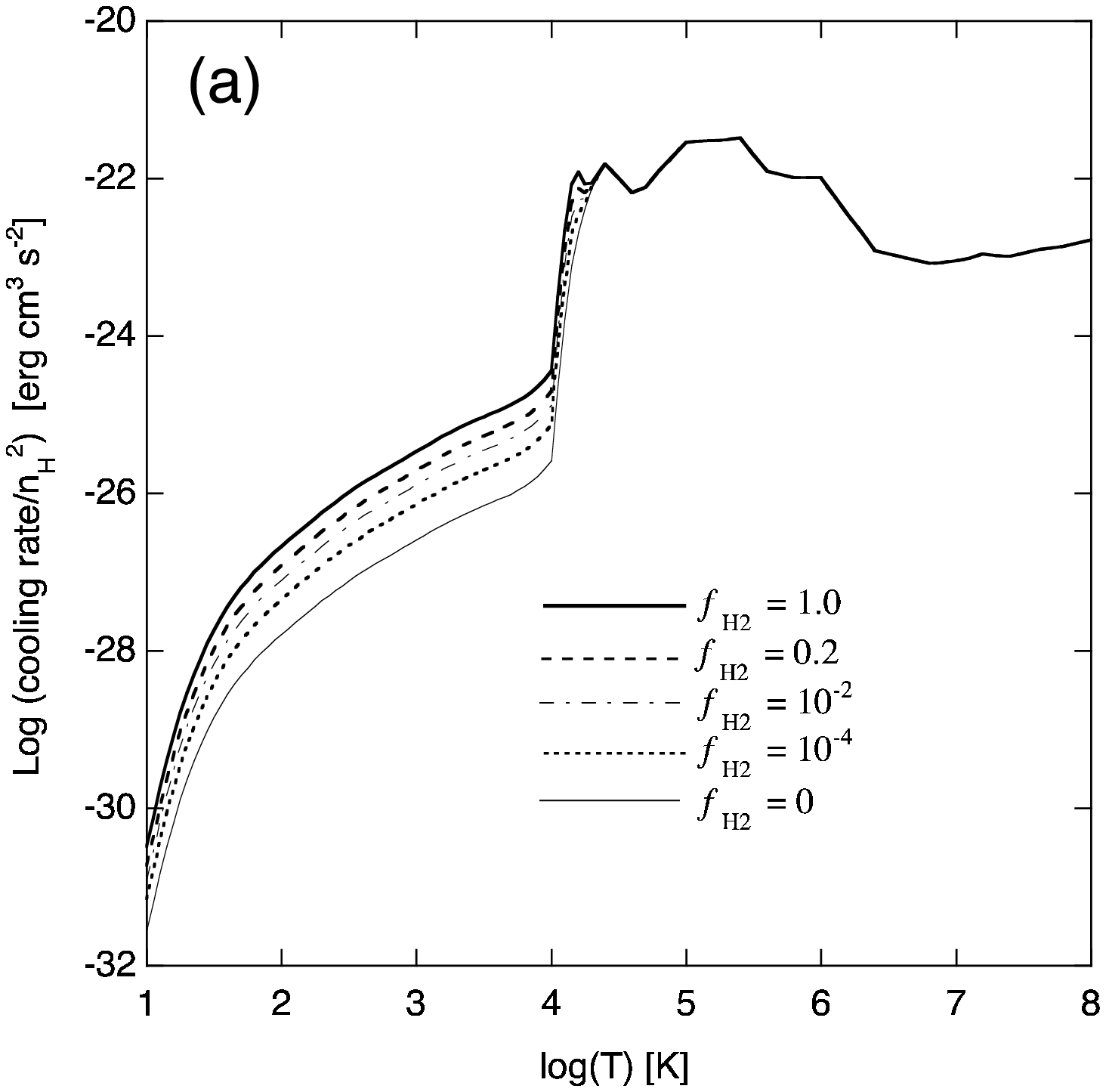} 
\includegraphics[width = 7.5cm]{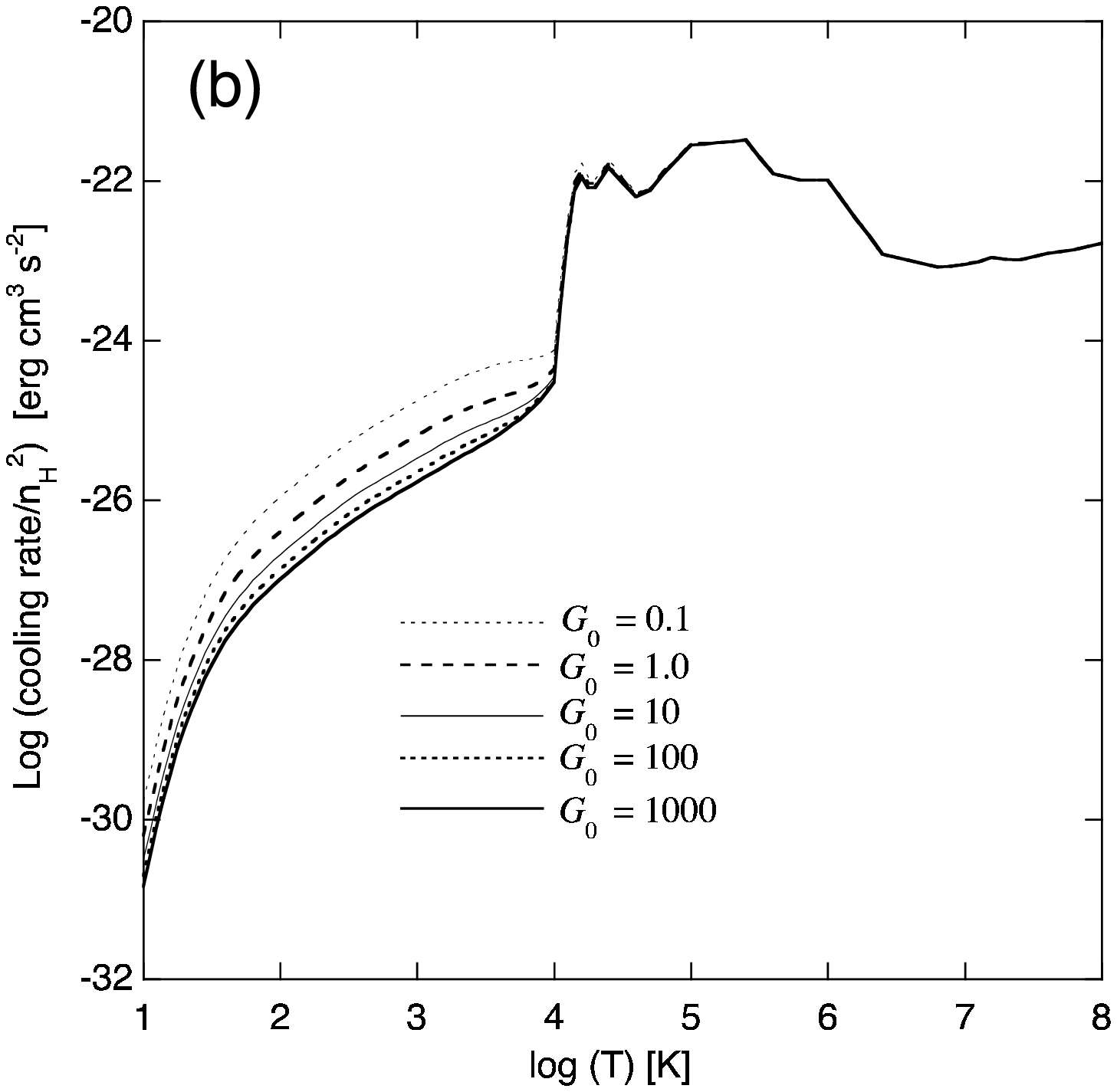} 
\caption{Cooling rate as a function of (a) $f_{\rm H_2}$ for $G_0 = 10$
 and (b) $G_0$ for  $f_{\rm H_2} = 1.0$}
\label{wada_fig: cooling1}
\end{figure}


\end{document}